\input harvmac
\def\O{{\cal O}}
\def\H{{\cal H}}
\def\l{\lambda}
\def\lp{\lambda^\prime}
\def\mnt{m_{\nu_\tau}}
\def\mnm{m_{\nu_\mu}}
\def\mne{m_{\nu_e}}
\def\mni{m_{\nu_i}}
\def\nt{{\nu_\tau}}
\def\nm{{\nu_\mu}}
\def\ne{{\nu_e}}
\def\sec{{\rm sec}}
\def\ra{\rightarrow}

\def\MNR{M^{\nu\ {\rm nr}}_{ij}}
\def\Mlo{M^{\nu\ {\rm loop}}_{ij}}

\def\eNR{\epsilon^{\rm nr}}
\def\elo{\epsilon^{\rm loop}}
\def\e{\epsilon}
\def\gsim{{~\raise.15em\hbox{$>$}\kern-.85em
          \lower.35em\hbox{$\sim$}~}}
\def\lsim{{~\raise.15em\hbox{$<$}\kern-.85em
          \lower.35em\hbox{$\sim$}~}}

\Title{hep-ph/9606251, WIS-96/21/May-PH}
{\vbox{\centerline{Neutrino Masses and Mixing in}
 \centerline{Supersymmetric Models without $R$ Parity}}}
\bigskip
\centerline{Francesca M. Borzumati, Yuval Grossman, Enrico Nardi
 and Yosef Nir}
\smallskip
\centerline{\it Department of Particle Physics}
\centerline{\it Weizmann Institute of Science, Rehovot 76100, Israel}
\bigskip
\baselineskip 18pt

\noindent
We study neutrino masses and mixing in Supersymmetric Models without
$R$ parity and with generic soft Supersymmetry breaking terms.
Neutrinos acquire mass from various sources: tree level
neutrino--neutralino mixing, loop effects and non--renormalizable
operators. Abelian horizontal symmetries (invoked to explain the
smallness and hierarchy in quark parameters) replace $R$ parity in
suppressing neutrino masses. We find lower bounds on the mixing angles:
$\sin\theta_{ij}\gsim m(\ell^-_i)/m(\ell_j^-)$ ($i<j$)
and unusual order of magnitude predictions for neutrino mass ratios:
$m(\ne)/m(\nm)\sim\sin^2\theta_{12}$; $m(\nu_i)/m(\nt)\sim 10^{-7}
\sin^2\theta_{i3}$ ($i=1,2$). Bounds from laboratory experiments exclude
$\mnt\gsim3\ MeV$ and  cosmological constraints exclude
$\mnt\gsim100\ eV$. Neither the solar nor the atmospheric
neutrino problems are likely to be solved by $\nm-\ne$ oscillations.
These conclusions can be evaded if holomorphy plays an
important role in the lepton Yukawa couplings.

\Date{5/96}

\newsec{Introduction}

The search for neutrino masses is one of the most promising directions to
find evidence for the incompleteness of the Standard Model.
Theoretical input is required in order to direct experiments to the
most plausible values of neutrino masses and mixing angles.
In particular, an understanding of the neutrino sector by the same
means that explain the quark and charged lepton parameters would be
desirable. Supersymmetry combined with horizontal symmetries can
provide this understanding
\nref\Ross{H. Dreiner {\it et al.}, Nucl. Phys. B436 (1995) 461.}%
\nref\RaSi{A. Rasin and J.P. Silva, Phys. Rev. D49 (1994) 20.}%
\nref\LMMM{Y. Grossman and Y. Nir, Nucl. Phys. B448 (1995) 30.}%
\nref\LLSV{G.K. Leontaris, S. Lola, C. Scheich, and J.D. Vergados,
hep-ph/9509351.}%
\nref\BLR{P. Binetruy, S. Lavignac, and P. Ramond, hep-ph/9601243.}%
\nref\ChLu{E.J. Chun and A. Lukas, hep-ph/9605377.}%
\refs{\Ross-\ChLu}.

In Supersymmetric models with the MSSM particle content and
with $R$ parity ($R_p$), lepton number
is violated by non--renormalizable terms only. Terms of the form
${1\over M}L\phi_u L\phi_u$ ($M$ is a high energy scale,
$L$ is the lepton doublet and $\phi_u$ is the $Y=+1/2$ Higgs doublet)
lead to neutrino masses of the see--saw type, $m_\nu\sim{\vev{\phi_u}^2
\over M}$. The consequences of Abelian horizontal symmetries in
this framework were investigated in ref. \LMMM.
A number of interesting order of magnitude relations among the
lepton parameters were found to hold in a large class of models:
\eqn\lmmma{{m_{\nu_i}\over m_{\nu_j}}\sim\sin^2\theta_{ij},}
\eqn\lmmmb{\sin\theta_{ij}\gsim{m_{\ell_i}\over m_{\ell_j}},}
\eqn\lmmmc{{m_{\nu_i}\over m_{\nu_j}}\gsim\left(
{m_{\ell_i}\over m_{\ell_j}}\right)^2,}
\eqn\lmmmd{m_\ne\lsim m_\nm\lsim m_\nt,}
where $i<j$ and $\ne,\nm,\nt$
denote the mass eigenstates with mixing of $\O(1)$ with $e,\mu,\tau$,
respectively. Interestingly, predictions analogous to \lmmmb\ and
\lmmmd\ apply to the quark sector (namely, $V_{ij}\gsim{m(u_i)\over
m(u_j)},\ {m(d_i)\over m(d_j)}$ and $V_{\rm CKM}\sim1$)
and are experimentally valid \LMMM.

The same horizontal symmetries that explain the smallness and hierarchy
in fermion parameters can naturally solve the problems related to
lepton flavor and lepton number violation that arise in Supersymmetric
models without $R_p$
\nref\HiKa{I. Hinchliffe and T. Kaeding, Phys. Rev. D47 (1993) 279.}%
\nref\BGNN{T. Banks, Y. Grossman, E. Nardi, and Y. Nir,
Phys. Rev. D52 (1995) 5319.}%
\nref\CHM{C.D. Carone, L.J. Hall, and H. Murayama, hep-ph/9512399;
hep-ph/9602364.}%
\refs{\HiKa-\CHM,\BLR}.\foot{
While horizontal symmetries can rather easily take
the role of $R_p$ in suppressing lepton number violation, it is much
more difficult to do so for baryon number violation
\ref\Hamo{V. Ben-Hamo and Y. Nir, Phys. Lett. B339 (1994) 77.}.
Therefore, as in ref. \BGNN, we simply assume that baryon number is a
symmetry of Nature.}
In this case, lepton number is violated by renormalizable terms, and the
resulting phenomenology is strikingly different from the one predicted
by Supersymmetric
$R_p$-symmetric models. In this work we examine the question of lepton
masses and mixing angles in models of Supersymmetry without $R_p$ but
with a horizontal symmetry.

\newsec{The Theoretical Framework}

We work in the framework of the Abelian horizontal symmetry $\H$
that has been introduced in refs.
\nref\LNS{M. Leurer, Y. Nir, and N. Seiberg,
 Nucl. Phys. B398 (1993) 319.}%
\nref\QSA{Y. Nir and N. Seiberg, Phys. Lett. B309 (1993) 337.}%
\nref\Seq{M. Leurer, Y. Nir, and N. Seiberg,
 Nucl. Phys. B420 (1994) 468.}%
\refs{\LNS-\Seq}.
$\H$ is explicitly broken by a small parameter $\l$ to which we
attribute charge --1 (and a numerical value of $\O(0.2)$, to explain
the Cabibbo angle). This can be viewed as the effective low energy theory
that comes from a Supersymmetric extension of the Froggatt--Nielsen
mechanism at a high scale
\ref\FrNi{C.D. Froggatt and H.B. Nielsen, Nucl. Phys. B147 (1979) 277.}.
Then, the following selection rules apply:
\item{(a)} Terms in the superpotential that carry charge $n\geq0$
under $\H$ are suppressed by $\O(\l^n)$, while those with $n<0$
are forbidden by holomorphy;
\item{(b)} Terms in the K\"ahler potential that carry charge $n$
under $\H$ are suppressed by $\O(\l^{|n|})$.

Without $R_p$ (or lepton number), there is a--priori no distinction
between the $Y=-1/2$ Higgs doublet $\phi_d$ and the three lepton
doublets $L_i$. (Wherever convenient, we denote the four doublets
by $L_\alpha$, $\alpha=0,1,2,3$.) The four doublets, however, carry
in general different  horizontal charges $H$. We identify the Higgs
doublet with
the doublet field that carries the smallest (positive) charge, which
we choose to be $L_0$ (we use interchangeably $L_0\equiv\phi_d$),
and we order the remaining doublets according to their charges:
\eqn\orderH{H(L_1)\geq H(L_2)\geq H(L_3)\geq H(\phi_d)\geq0\,.}
A similar ordering is made for the three generations of
$\bar\ell_i$ (charged lepton singlets), $Q_i$ (quark doublets), and
$\bar d_i$ (down quark singlets).
The model is phenomenologically viable if (for $\tan\beta\sim1$)
the following condition holds~\BGNN:
\eqn\nutaubound{H(L_i)-H(\phi_d)\geq3.}

Our methods of analyzing lepton and neutralino mass matrices
are described in detail in refs. \LMMM\ and \BGNN, respectively.
Specifically, we use the following selection rules
to estimate the magnitude of the various contributions.
For the quadratic terms in the superpotential,
$\mu_\alpha L_\alpha\phi_u$, it is :
\eqn\selectm{\mu_\alpha\sim\cases{
\tilde\mu\l^{H(L_\alpha)+H(\phi_u)}&$H(L_\alpha)+H(\phi_u)\geq0$,\cr
\tilde m\l^{|H(L_\alpha)+H(\phi_u)|}&$H(L_\alpha)+H(\phi_u)<0$.\cr}}
Here $\tilde\mu$ is the natural scale for the $\mu$ terms and
$\tilde m$ is the Supersymmetry breaking scale. For simplicity we assume
that $\tilde\mu$ is $\O(\tilde m)$. As $\mu_0$ is phenomenologically
required to be of $\O(\tilde m)$, we take $H(\phi_d)+H(\phi_u)\sim0$.
Modifications to the case where the natural scale for $\mu$ is, say,
$M_{\rm Planck}$ and it is suppressed down to $\tilde m$ by the
horizontal symmetry, as in the models of \Seq\ and \BGNN, are
straightforward. The selection rule for the coupling of the quadratic
soft Supersymmetry breaking terms, $B_\alpha L_\alpha\phi_u$
(here $L_\alpha$ stand for the scalar components) is:
\eqn\selectB{B_\alpha\sim\tilde m^2\l^{|H(L_\alpha)+H(\phi_u)|}.}
Finally, the selection rules for the trilinear terms
$\lp_{\alpha jk}L_\alpha Q_j\bar d_k$ and
$\l_{\alpha\beta k}L_\alpha L_\beta\bar\ell_k$ are
\eqn\selectlp{\lp_{\alpha jk}\sim\cases{
\l^{H(L_\alpha)+H(Q_j)+H(\bar d_k)}&$H(L_\alpha)+H(Q_j)+H(\bar d_k)
\geq0$,\cr   0&$H(L_\alpha)+H(Q_j)+H(\bar d_k)<0$,\cr}}
\eqn\selectl{\l_{\alpha\beta k}\sim\cases{
\l^{H(L_\alpha)+H(L_\beta)+H(\bar\ell_k)}&
$H(L_\alpha)+H(L_\beta)+H(\bar\ell_k)\geq0$,\cr
0&$H(L_\alpha)+H(L_\beta)+H(\bar\ell_k)<0$.\cr}}
Note that $\lp_{0jk}$ and $\l_{0jk}$ are practically the Yukawa couplings
for the down sector and for the charged lepton sector.

The order of magnitude relations \lmmma--\lmmmd\ were derived in a large
class of models where all entries in the lepton mass matrices
carry positive charges. (They are actually applicable in a larger class
of models, where the holomorphy--induced zero entries, if any,
do not affect the physical parameters.) In this work we restrict
ourselves to this class of models.

\newsec{Neutrino Masses and Mixing}

There are several important sources for neutrino masses in this
framework, each giving a different scale:
renormalizable tree--level mixing with neutralinos
\nref\AuMo{C.S. Aulakh and R.N. Mohapatra, Phys. Lett. 119B (1982) 136.}%
\nref\HaSu{L.J. Hall and M. Suzuki, Nucl. Phys. B231 (1984) 419.}%
\nref\LHLee{L.-H. Lee, Phys. Lett. 138B (1984) 121;
Nucl. Phys. B246 (1984) 120.}%
\nref\RoVa{G.G. Ross and J.W.F. Valle, Phys. Lett. 151B (1985) 375.}%
\nref\Elli{J. Ellis {\it et al.}, Phys. Lett. 150B (1985) 142.}%
\nref\Daws{S. Dawson, Nucl. Phys. B261 (1985) 297.}%
\nref\SaVa{A. Santamaria and J.W.F. Valle, Phys. Lett. B195 (1987) 423.}%
\nref\BHH{D.E. Brahm, L.J. Hall, and S. Hsu,
 Phys. Rev. D42 (1990) 1860.}%
\nref\RoVa{J.C. Romao and J.W.F. Valle, Nucl. Phys. B381 (1992) 87.}%
\nref\JoNo{A.S. Joshipura and M. Nowakowski,
 Phys. Rev. D51 (1995) 2421.}%
\refs{\AuMo-\JoNo}; quark--squark and lepton--slepton loop corrections
\nref\DiHa{S. Dimopoulos and L.J. Hall, Phys. Lett. B207 (1987) 210.}%
\nref\BaMo{K.S. Babu and R.N. Mohapatra,
 Phys. Rev. Lett. 64 (1990) 1705.}%
\nref\BGMT{R. Barbieri, M.M. Guzzo, A. Masiero, and D. Tommasini,
 Phys. Lett. B252 (1990) 251.}%
\nref\RoTo{E. Roulet and D. Tommasini, Phys. Lett. B256 (1991) 218.}%
\nref\EMR{K. Enkvist, A. Masiero, and A. Riotto,
 Nucl. Phys. B373 (1992) 95.}%
\nref\GRT{R.M. Godbole, P. Roy, and X. Tata,
 Nucl. Phys. B401 (1993) 67.}%
\nref\Hemp{R. Hempfling, hep-ph/9511288.}%
\nref\CaWh{B. de Carlos and P.L. White, hep-ph/9602381.}%
\refs{\HaSu,\DiHa-\CaWh}; and non--renormalizable see--saw contributions
\nref\Yana{T. Yanagida, {\it in} Proc. Workshop on Unified theory
and baryon number in the universe, eds. O. Sawada and A. Sugamoto
(KEK, 1979).}%
\nref\GRS{M. Gell-Mann, P. Ramond and R. Slansky, {\it in} Supergravity,
eds. P. van Nieuwenhuizen and D. Freedman (North-Holland, 1980).}%
\refs{\Yana-\GRS}. We now discuss each contribution in turn.
In our various estimates we take $\tan\beta\sim1$.

\item{(i)} {\it Renormalizable tree--level contributions.}

\noindent
These contributions arise when the $\mu$--terms in the superpotential
and the Supersymmetry breaking $B$ terms in the scalar potential are
misaligned, $B_\alpha\neq B \mu_\alpha\,$, or the Supersymmetry
breaking scalar masses do not satisfy the eigenvalue condition
$m^2_{\alpha\beta} \mu_\beta = {\tilde m}^2 \mu_\alpha$~\BGNN.
This yields  misalignment between the VEVs
$v_\alpha \equiv \langle L_\alpha\rangle$ and the $\mu_\alpha$ terms,
\eqn\mis{\quad\qquad\qquad\qquad\quad\sin^2\xi =
{1\over 2}{\sum_{\alpha,\beta}(\mu_\alpha v_\beta -\mu_\beta v_\alpha)^2
 \over \mu^2 v_d^2} \,, \qquad (\mu^2\equiv\mu_\alpha\mu_\alpha\,,
 \ v_d^2 \equiv v_\alpha v_\alpha)}
which induces neutrino mixing with the neutralinos. Only one neutrino
acquires a mass from this effect: $m_{\nu} \sim m_Z \, \sin^2\xi\,$
(here $m_Z$ stands for the electroweak or Supersymmetry breaking scale).
In the absence of any symmetry reason for alignment, we expect
$\sin\xi\sim1$ and the natural scale for $m_\nu$ is  the electroweak
scale. The further required suppression comes from $\H$-violation,
$v_i/v_0 \sim \l^{[H(L_i)-H(\phi_d)]}\,$ \BGNN. Thus,
\mis\ together with \orderH\  gives
\eqn\mnutau{\mnt\sim\l^{2[H(L_3)-H(\phi_d)]}m_Z.}
The massive neutrino is then $\nu_\tau\,$, which is close to the
interaction eigenstate with the smallest horizontal charge among the
$L_i$.  The experimental upper bound $\mnt\leq24\ MeV$
\ref\Alep{D. Buskulic {\it et al.}, ALEPH Collaboration,
Phys. Lett. B349 (1995) 585.}, when
confronted with \mnutau, is the source of the constraint \nutaubound.

\item{(ii)} {\it Quark--squark loop contributions.}

\noindent
Loops with down quark and squarks contribute
\eqn\mnuBGN{{1\over 2} \,\Mlo\sim{3\lp_{ikl}\lp_{jmn}\over16\pi^2}
{(M^d)_{lm}(\tilde M^{d2}_{LR})_{kn}\over\tilde m^2},}
where $M^d$ is the $d$--quark mass matrix, and $\tilde M^{d2}_{LR}$ is
the left--right sector in the $\tilde d$-squark mass-squared matrix.
The experimental value of $V_{cb}\sim m_s/m_b$
strongly suggests that $H(\bar d_2)\!=\!H(\bar d_3)$ and consequently
$(M^d)_{32}\sim(M^d)_{33}\sim m_b$ and $(\tilde M^{d2}_{LR})_{32}\sim
(\tilde M^{d2}_{LR})_{33}\sim\tilde m m_b$. From this, together with
\selectlp, we learn that the largest contributions to \mnuBGN\
come from $k=m=3$, and $l\,,n=2,3$.
This, in general, gives mass to the two light neutrinos.
Taking into account that $\lp_{033}\sim m_b/m_Z$, one obtains:
\eqn\ratLO{\eqalign{
{\mnm^{\rm loop}\over\mnt}\sim&\ \elo\l^{2[H(L_2)-H(L_3)]},\cr
{\mne^{\rm loop}\over\mnt}\sim&\ \elo\l^{2[H(L_1)-H(L_3)]},\cr
\elo=&\ {3m_b^4\over 8 \pi^2 m_Z^4}\sim10^{-7},\cr}}
which implies the upper limit for the loop--induced masses
$m_{\nu_i}\lsim10^{-7}\mnt\lsim1~eV$,~$i=1,2$.

Note that if the Supersymmetry breaking trilinear scalar couplings
are proportional to the Yukawa couplings
($\tilde M^{d2}_{LR} = {\tilde m} M^d$), the dominant
quark--squark loop contributions to  \mnuBGN\  yield a degeneracy
in the mass matrix, and only one neutrino in  \ratLO\
acquires mass.
In the presence of a tree--level mass \mnutau\ for $\nt\,$,
this mass eigenstate is $\nm$ (and is close to the interaction
eigenstate $L_2$). The same result applies also to the case that
$(M^d)_{32}\ll(M^d)_{33}$ and $(\tilde M^{d2}_{LR})_{32}\ll
(\tilde M^{d2}_{LR})_{33}$. Then a contribution to $\mne$ arises
from quark-squark loops with e.g.
$k\!=\!n\!=\!2$, $l\!=\!m\!=\!3$, giving $\mne/\mnt\sim
{3m_s^2m_b^2\over8\pi^2 m_Z^4}\l^{2[H(L_1)-H(L_3)]}$, about two to three
orders of magnitude below \ratLO. This is somewhat smaller than the
contribution from lepton--slepton loops discussed below.

\item{(iii)} {\it Lepton--slepton loop contributions.}

\noindent
Loops with charged leptons and sleptons contribute
\eqn\mnuBGNl{{1\over 2}\, \Mlo\sim{\l_{ikl}\l_{jmn}\over16\pi^2}
{(M^\ell)_{lm}(\tilde M^{\ell2}_{LR})_{kn}\over\tilde m^2}.}
Using \selectl, we learn that the largest contribution from
\mnuBGNl\ has
\eqn\ratLOl{\elo={m_\tau^4\over 8 \pi^2 m_Z^4}\sim10^{-9},}
about two orders of magnitude lower than the dominant quark--squark
contributions. As already mentioned, this contribution plays a
significant role only when $\tilde M^{d2}_{LR} \simeq {\tilde m} M^d$
holds to a good approximation.

\item{(iv)} {\it Non--renormalizable contributions.}

\noindent
The dimension--5 terms ${1\over M}\nu_i\nu_j\phi_u\phi_u$ give \LMMM:
\eqn\mtLMMM{{1\over 2}\,\MNR\sim\l^{H(L_i)+H(L_j)+2H(\phi_u)}
{m_Z^2\over M}.}
This, in general, contributes to both light neutrinos:
\eqn\ratNR{\eqalign{
{\mnm^{\rm nr}\over\mnt}\sim&\ \eNR\l^{2[H(L_2)-H(L_3)]},\cr
{\mne^{\rm nr}\over\mnt}\sim&\ \eNR\l^{2[H(L_1)-H(L_3)]},\cr
\eNR=&\ \l^{2[H(\phi_u)+H(\phi_d)]}{m_Z\over M}\sim
10^{-7}\ \left({10^{9}\ GeV\over M}\right).\cr}}

The relative importance of the non--renormalizable and loop contributions
to $\mnm$ and $\mne$ depends on the scale $M$ (which is, roughly
speaking, the natural scale for the masses of right--handed neutrinos).
For $M\gsim10^{9}\ GeV$, the leading contributions come from loops, while
for $M\lsim10^{9}\ GeV$, the non--renormalizable contributions dominate.

Adding up the various contributions, and defining
\eqn\defeps{\e={\rm max}(\eNR,\elo),}
leads to the following order of magnitude
estimates for the neutrino masses and mixing angles:
\eqn\masses{\eqalign{
\mnt/m_Z\sim&\ \l^{2[H(L_3)-H(\phi_d)]},\cr
\mni/\mnt\sim&\ \e\l^{2[H(L_i)-H(L_3)]}\ \qquad  (i=1,2),\cr}}
\eqn\mixing{\sin\theta_{ij}\sim \l^{H(L_i)-H(L_j)}\ \ \ \qquad \quad (i<j).}
The charged lepton mass ratios are estimated to be
\eqn\mcharged{m_{\ell_i}/m_{\ell_j}\sim\l^{H(L_i)+H(\bar\ell_i)
-H(L_j)-H(\bar\ell_j)}\,.}

The charged current mixing matrix mixes not only the leptons
among themselves, but also leptons with higgsinos and gauginos:
\eqn\lepzH{\eqalign{
\sin\theta_{\nu_i\tilde\phi_d^-}\sim&\ \lambda^{H(L_i)-H(\phi_d)},\cr
\sin\theta_{\nu_i\tilde w^-}\sim&\ \lambda^{H(L_i)+H(\phi_u)}.\cr}}
The fact that the neutrino mass eigenstates have an isotriplet
$\tilde w_3$ component in them, leads to flavor changing couplings of the
$Z$-boson to neutrinos $\sim g\Omega_{ij}Z\nu_i\bar\nu_j$:
\eqn\Oij{\Omega_{ij}\sim\l^{H(L_i)+H(L_j)+2H(\phi_u)}.}

The estimates \masses\ and \mixing\ give the following relations between
neutrino masses and mixing angles:
\eqn\masmix{\eqalign{
{\mnm\over\mnt}\sim&\ \e\sin^2\theta_{23},\cr
{\mne\over\mnt}\sim&\ \e\sin^2\theta_{13},\cr
{\mne\over\mnm}\sim&\ \sin^2\theta_{12}.\cr}}
There are two order of magnitude relations that are
independent of $\e\>$:
\eqn\predict{\sqrt{\mne\over\mnm}\sim
\sin\theta_{12}\sim{\sin\theta_{13}\over\sin\theta_{23}}\,,}
so that for the two light neutrinos \lmmma\ still holds.
The order of magnitude inequality \lmmmb\ is maintained:
\eqn\bgnnb{\sin\theta_{ij}\gsim{m_{\ell_i}\over m_{\ell_j}}\ \qquad \ (i<j).}
The relation \lmmmd\ is also maintained,
\eqn\bgnnc{m_\ne\lsim m_\nm\lsim m_\nt.}
However, unlike \lmmmc, the light neutrinos are much lighter than $\nt$:
\eqn\upper{{m_{\nu_i}\over\mnt}\lsim\ \e\ \qquad \ (i=1,2).}
For a scale $M\gsim10^{9}\ GeV$, this gives ${\mni\over\mnt}\lsim
10^{-7}$. It is interesting that these models can naturally give mixing
angles of $\O(1)$ with the third generation \JoNo\ while the
corresponding mass ratios are very small. (For different mechanisms that
give such a result, see
\nref\NRT{E. Nardi, E. Roulet, and D. Tommasini,
 Phys. Lett. B327 (1994) 319.}%
\nref\BaBa{K.S. Babu and S.M. Barr, hep-ph/9511446.}%
\refs{\NRT-\BaBa}.)

\newsec{Theory Confronts Experiment}

In this and the next sections we show that the order of magnitude
relations derived in the last section, when combined with various
experimental and cosmological constraints, exclude large regions
of the mass--mixing parameter space. Most of our discussion in these two
sections is independent of the question of $R_p$ violation.

As the charged lepton masses are known, eq. \bgnnb\ provides
significant lower bounds on the lepton mixing angles. With
$m_e/m_\mu\sim\l^3$ and $m_\mu/m_\tau\sim\l^2$, we get
\eqn\bnumb{\sin\theta_{23}\gsim\l^2,\ \ \
\sin\theta_{13}\gsim\l^5,\ \ \ \sin\theta_{12}\gsim\l^3.}

The lower bound on $\sin\theta_{23}$ is particularly significant.
First, if $\mnt$ is in the appropriate range, $\nm-\nt$
oscillations will be observed in the CHORUS, NOMAD and E803 experiments.
Second, combining it with the upper bound $\sin^2\theta_{23}=
BR(\pi\ra\mu\nt)\leq6\times10^{-5}$ for $\mnt\gsim3\ MeV$
\ref\Daum{M. Daum {\it et al.}, Phys. Rev. D36 (1987) 2624.}\ results in
\eqn\pidecay{\mnt\lsim3\ MeV.}
Third, in combination with the bound on $\nm-\nt$ oscillations,
$\sin^22\theta_{23}\leq0.004$ for $\Delta m^2\gsim100\ eV^2$
\ref\Ushi{N. Ushida {\it et al.}, Fermilab E531 Collaboration,
Phys. Rev. Lett. 57 (1986) 2897.},
it gives\foot{The interpretation of oscillation experiments might change
in the framework of Supersymmetry without $R_p$ because new neutrino
interactions are introduced
\ref\Gros{Y. Grossman, Phys. Lett. B359 (1995) 141.}. In our case,
however, the horizontal suppression makes these new interactions
practically negligible.}
\eqn\mtosc{\sin\theta_{23}\sim\l^2\ \ \ {\rm for}\ \ \
10\ eV\lsim\mnt\lsim3\ MeV.}
As we predict $\sin\theta_{13}\lsim\sin\theta_{23}$,
\mtosc\ implies also
\eqn\etosc{\sin\theta_{13}\lsim\l^2\ \ \ {\rm for}\ \ \
10\ eV\lsim\mnt\lsim3\ MeV.}
This bound is stronger than the bound from $\ne-\nt$ oscillations,
$\sin^22\theta_{13}\leq0.12$ for $\Delta m^2\gsim100\ eV^2$
\Ushi, which,  in this range, gives $\sin\theta_{13}\lsim\l$. The latter
bound, however, holds independently of whether holomorphy plays a role
in determining the mixing angles.

Eqs. \pidecay, \mtosc\ and \etosc\
are applicable also in models with $R_p$ because they result
from \bgnnb\ which holds independently of $R_p$ violation.

\newsec{Theory Confronts Cosmology}

Cosmological considerations related to the age and the present
energy density of the Universe provide a constraint on the mass
and lifetime of neutrinos. For masses in the range $100\ eV$ --
a few $MeV$, the constraint reads (see e.g.
\ref\HaNi{H. Harari and Y. Nir, Nucl. Phys. B292 (1987) 251.})
\eqn\cosmo{\mnt^2\tau_\nt\lsim2\times10^{20}\ eV^2\ \sec.}
The framework of Abelian horizontal symmetries allows
an estimate of the neutrino decay rates. It is interesting to find
whether $\nt$ could have a fast enough decay mode
to fulfill \cosmo\ and have its mass above $100\ eV$.

The dominant decay modes are most likely those
which proceed via gauge interactions. The bound \pidecay\ leaves
only a very small window where the $W$--mediated tree level
$\nt\ra e^+e^-\ne$ is allowed. The rate can be estimated to be:
\eqn\numbertree{\eqalign{&
{\Gamma(\nt\ra e^+e^-\ne)\over \Gamma(\tau\ra e\bar\ne\nt)}=
{\mnt^5\over m^5_\tau}\sin^2\theta_{13}\cr \Longrightarrow&\
\mnt^5\tau_{\nt}\sim
\left(\l^5\over\sin\theta_{13}\right)^2\ 3\times10^{41}\ eV^5\ \sec.\cr}}
Together with \pidecay, we find that \cosmo\ is satisfied only for
$\sin\theta_{13}\gsim\l^4$. Since there are charged
particles in the final state, however,
a stronger bound (from considerations of the
cosmic microwave background radiation) applies,
$\tau_\nt\lsim10^4\ \sec$. This cannot be satisfied for
$\mnt\lsim3\ MeV$ and $\sin\theta_{13}\lsim\l^2$.
Furthermore, detailed studies of the effects  of
a massive $\nt$ during the Nucleosynthesis era
\nref\KTCS{E. Kolb, M.S. Turner, A. Chakravorty, and D.N. Schramm,
 Phys. Rev. Lett. 67 (1991) 533.}%
\nref\DoRo{A.D. Dolgov and I.Z. Rothstein,
 Phys. Rev. Lett. 71 (1993) 476.}%
\nref\Kawa{M. Kawasaki {\it et.al.}, Nucl. Phys. B 419 (1994) 105.} %
\refs{\KTCS-\Kawa}
suggest that for $\mnt\gsim0.5\ MeV$, and independently of the decay
modes, $\tau_\nt\lsim10^2\ \sec$ is required, which closes the window
even more firmly. Therefore we conclude that $\nt\ra e^+e^-\ne$ does
not open any window for a heavy $\nt$.
Again, this conclusion holds also for models with $R_p$.

All other decay modes are flavor changing neutral current processes.
There are three types of contributions to such processes:
\item{(a)} Loop diagrams with gauge particles, suppressed by
the charged current mixing angles;
\item{(b)} Tree level $Z$--mediated decays
suppressed by the $\Omega_{ij}$ mixing angles;
\item{(c)} Tree level slepton--mediated decays suppressed
by the selection rules for the $\l_{ijk}$ couplings.

The first class is common to models with and without $R_p$,
but the other two are present only in  $R_p$--violating models.
In any case, we found that none of these channels is fast
enough to allow $\mnt\gsim100\ eV$. For example,
the rate for the $Z$--mediated $\nt\ra3\nm$ can be estimated to be:
\eqn\Zmediate{\eqalign{
&{\Gamma(\nt\ra3\nm)\over\Gamma(\tau\ra e\bar\ne\nt)}\sim
{\mnt^5\over m^5_\tau}\Omega^2_{23}\cr \Longrightarrow&\
\mnt^5\tau_{\nt}\sim\left({\l^6\over\Omega_{23}}
\right)^2 10^{43}\ eV^5\ \sec.\cr}}
This is significantly suppressed compared to \numbertree\
and does not satisfy \cosmo.

As $\nt$ is predicted to be the heaviest among the neutrinos,
we conclude that in the framework of Supersymmetry and Abelian
horizontal symmetry with or without $R_p$ (and assuming that holomorphy
does not play a role in determining $\sin\theta_{23}$)
\eqn\upper{m_{\nu_i}\lsim100\ eV}
holds for all neutrino masses.
We note that in the framework of a single $U(1)$ or $Z_n$
broken by $\l\sim0.2$, this requires $H(L_i)-H(\phi_d)\gsim6$
which may be too large for reasonable models. In some
models of ref. \LNS, however, where the symmetry breaking parameters
are much smaller, this can be achieved with charge differences $\leq2$.

\newsec{Solar and Atmospheric Neutrinos}

The upper bound $\mnt\lsim100\ eV$ leads to even stronger bounds
on $\mnm$ and $\mne$. These bounds, however, depend on $M$
and on $\tan\beta$.
We believe that the most likely situation is (a) $M\gsim10^{9}\ GeV$
(which, for example, applies in all models where the Supersymmetric
extension of the Standard Model is valid up to some GUT scale)
and (b) $\tan\beta\sim1$ (which is the {\it natural} value
\nref\NeRa{A.E. Nelson and L. Randall, Phys. Lett. B316 (1993) 516.}%
\nref\RaSa{R. Rattazzi and U. Sarid, Phys. Rev. D53 (1996) 1553.}%
\refs{\NeRa-\RaSa}). Then in \defeps\
$\e\sim10^{-7}$ leads to
\eqn\upperer{\mnt\lsim100\ eV,\ \ \ \mne\lsim\mnm\lsim10^{-5}\ eV.}
This bears important consequences for the solar and atmospheric
neutrino problems. The value $\mnm\lsim10^{-5}\ eV$ is inconsistent with
\eqn\MSW{m_\nm^2-m_\ne^2\sim6\times10^{-6}\ eV^2}
that is required to solve the solar neutrino problem through
the MSW mechanism  (see e.g.
\ref\HaLa{N. Hata and P. Langacker, Phys. Rev. D50 (1994) 632.}),
but is consistent with $m_\nm^2-m_\ne^2\sim10^{-11}\ eV^2$
that could solve it through vacuum oscillations (see e.g.
\ref\CFFL{E. Calabresu, N. Ferrari, G. Fiorentini, and M. Lissia,
 Astropart. Phys. 4 (1995) 159.}).
It  is also inconsistent with
\eqn\atmo{\mnm^2-\mne^2\sim10^{-2}\ eV^2}
that is required to solve the atmospheric neutrino problem through
$\nm-\ne$ oscillations (see e.g.
\ref\FoLi{G.L. Fogli and E. Lisi, Phys. Rev. D52 (1995) 2775.}).

The solar neutrino problem can still be solved by the MSW mechanism with
$\nt-\ne$ oscillations (if $\mnt\sim10^{-3}\ eV$) or the atmospheric
neutrino problem can be explained by $\nt-\nm$ oscillations
(if $\mnt\sim10^{-1}\ eV$). However, we find that:
\item{a.} The two problems cannot be solved simultaneously;
\item{b.} The required horizontal charges are inconveniently large:
$H(L_3)-H(\phi_d)\sim10$ and $H(L_1)-H(\phi_d)\sim12$ to get
the small angle MSW solution for the solar neutrino problem, and
$H(L_{2,3})-H(\phi_d)\sim8$ to solve the atmospheric neutrino problem;
\item{c.} Such a light $\nt$ does not contribute to the dark matter
and cannot play any role in structure formation. This would require
\eqn\dark{\mnt\sim10\ eV.}

This situation is very different from the Supersymmetric
models with $R_p$, where \dark\ and \MSW\ can be simultaneously
accommodated \LMMM.

Things are different if we relax either of our two extra assumptions.
As the tree level contribution to $\mnt$ is suppressed by $\tan^2\beta$
and the loop contributions are enhanced by $\tan^3\beta$, a large
$\tan\beta$ would give a large $\e$ ({\it e.g.} $\tan\beta\sim20$ gives
$\e\sim10^{-2}$). Then we can easily accommodate the dark matter
\dark\ and solar neutrino \MSW\ constraints
($H(L_2)=H(L_3)+1=H(\phi_d)+6$). Alternatively, the solar and atmospheric
neutrino problems can be solved simultaneously
($H(L_3)=H(L_2)=H(\phi_d)+6$). It is non--trivial that, in this scenario,
$H(L_2)$ and $H(L_3)$ which
are fixed by the requirements on $m_\nm$ and $\mnt$ give, at the
same time, $\sin\theta_{23}=\O(1)$ as required to solve the
atmospheric neutrino problem.

For $M<10^9\ GeV$, the non--renormalizable contributions to $\mnm$
dominate over the loop corrections.
In this case $\mnt$ can account for the dark matter, while
$m_{\nm}$  can accommodate the solar neutrino
constraint \MSW\ (however, this requires $M\lsim10^6\ GeV$).

Finally, if holomorphy does play an important role in the physical
parameters, then even the conclusions of sections 3--5 can
be evaded. For example, we can construct models where $\sin\theta_{23}
\ll\l^2$, which would allow for $\mnt$ above the $3\ MeV$ bound of section 4.
Then, with $\sin\theta_{13}\sim\l^4$ (which is marginally compatible with
the experimental limits
\nref\DLR{De Leener-Rosier {\it et al.}, Phys. Rev. D43 (1991) 3611.}%
\nref\Zub{K. Zuber, hep-ph/9605405.}%
\refs{\DLR-\Zub}), the decay
$\nt\ra e^+e^-\ne$ can still open a window for $\mnt$ close to its
experimental bound. Explicit examples of models where holomorphy
induces approximate zeros in the mass matrices and affects physical
parameters can be found in refs. \refs{\QSA,\Seq,\LMMM}.

\newsec{Discussion}

Models of Supersymmetry with Abelian horizontal symmetries have
interesting implications for neutrino masses and mixing. We distinguish
three cases:
\item{(i)} Models with $R_p$.
\item{(ii)} Models without $R_p$ and with generic
soft Supersymmetry breaking terms
(up to the selection rules from the horizontal symmetry).
\item{(iii)} Models without $R_p$ but with universal Supersymmetry
breaking terms implying alignment at a high scale (namely $B_\alpha
=B\mu_\alpha$ and $m^2_{\alpha\beta}=\tilde m^2\delta_{\alpha\beta}$).

Class (i) was analyzed in \LMMM. Class (ii) has been studied in this
work. Class (iii), which yields a scenario quite different
from the one investigated here, will be discussed in a forthcoming paper.
We now compare the predictions of class (ii) with those of (i).

In a large class of models, where holomorphy does not introduce
zeros in the mass matrices (or, if there are such zeros, they
do not affect the order of magnitude of the physical parameters),
we find the following order of magnitude relations among the mass ratios
and mixing angles:
\eqn\conRp{\matrix{
\qquad &\qquad \mnm/\mnt\qquad &\qquad\mne/\mnt\qquad
&\qquad\mne/\mnm\qquad  &\qquad\qquad\cr &&&&\cr
(i)&\sin^2\theta_{23}&\sin^2\theta_{13}&\sin^2\theta_{12}&
\qquad\qquad \cr &&&&\cr
(ii)&10^{-7}\sin^2\theta_{23}&10^{-7}\sin^2\theta_{13}&
\sin^2\theta_{12}&\qquad\qquad\cr}}
Note that the following
relation among the mixing angles holds in both classes:
\eqn\general{(i)\,,\>(ii)\,:
\qquad\qquad \sin\theta_{13}\sim\sin\theta_{12}\,\sin\theta_{23}. \
\quad\quad\qquad\qquad\qquad\qquad\qquad }
Furthermore, since $\sin\theta_{ij}\gsim m_{\ell_i}/m_{\ell_j}$ holds
independently of $R_p$,
\eqn\lowsin{(i)\,,\>(ii)\,:
\qquad\qquad \sin^2\theta_{23}\gsim10^{-3},\ \ \ \sin^2\theta_{13}\gsim
10^{-7},\ \ \ \sin^2\theta_{12}\gsim10^{-4}.   }
This leads to the following ranges for the mass ratios:
\eqn\conRpp{\matrix{\qquad&\qquad\mnm/\mnt\qquad&\qquad\mne/\mnt\qquad
&\qquad\mne/\mnm\qquad&\qquad\qquad\cr  &&&&\cr
(i)&10^{-3}\>-\>1\quad\ \ &10^{-7}\>-\>1\quad\ \ &10^{-4}\>-\>1\quad\ \
 &\qquad\qquad\cr &&&&\cr
(ii)&10^{-10}\>-\>10^{-7}&10^{-14}\>-\>10^{-7}&10^{-4}\>-\>1\quad\ \
 &\qquad\qquad\cr}}
We conclude that measurements of the lepton mixing angles would
test the Supersymmetric Abelian horizontal symmetry framework
while measurements of neutrino mass ratios will serve to distinguish
between models with or without $R_p$.

In ref. \HaNi, it was shown that $\mnm/\mnt\gsim(m_\mu/m_\tau)^2$
together with cosmological considerations, strongly suggests that all
neutrinos are lighter than $\O(100\ eV)$. Models without $R_p$
predict $m_\nm/m_\nt\ll(m_\mu/m_\tau)^2$ but we still find that
all neutrinos are lighter than $\O(100\ eV)$. This is a consequence
of the fact that there is no decay mode large enough
to fulfill the cosmological constraints on massive neutrinos.

In models with $R_p$, one can accommodate $\mnt$ to contribute
sizeably to the cosmological  dark matter, as well as
$\mnm$ in the correct range  required by the  MSW
solution of the solar neutrino problem.  In models without $R_p$
(and without any alignment condition)
we find that, unless the scale of New Physics $M$ is surprisingly
low ($\lsim10^6\ GeV$), $\nm$ is too light to
play any role for matter enhanced oscillations of the solar $\ne$'s.

Finally, we emphasize that our various predictions are not
entirely generic to models of Abelian horizontal symmetries.
As described briefly in section 6, a large $\tan\beta$ and/or
a small scale $M$ would modify our discussion of the
solar (and atmospheric) neutrino problem. But more important, one can
construct models where holomorphy plays an important role
and circumvents the otherwise model--independent predictions
of eqs. \predict, \bgnnb\ and \bgnnc.


\vskip 1cm
{\bf Acknowledgments:}
FMB acknowledges discussions with R. Hempfling.
YN is supported in part by the United States -- Israel Binational
Science Foundation (BSF), by the Israel Commission for Basic Research,
and by the Minerva Foundation (Munich).

\listrefs
\end